\documentclass[11pt,twoside]{book}
\usepackage{konkolyproc2}
\usepackage{longtable}
\usepackage{amsmath,amssymb}
\usepackage{graphicx}
\usepackage{lscape}
\usepackage{index}
\usepackage{natbib}
\usepackage{bigdelim}
\usepackage{multirow}
\makeindex

\begin{document}

\pagestyle{myheadings}
\setcounter{equation}{0}\setcounter{figure}{0}\setcounter{footnote}{0}\setcounter{section}{0}\setcounter{table}{0}\setcounter{page}{1}
\markboth{Smolec}{RRL2015 Conf. Papers}
\title{On period ratios in modulated double-mode RR Lyrae stars}
\author{Rados\l{}aw Smolec}
\affil{Nicolaus Copernicus Astronomical Center, PAS, Warsaw, Poland}

\begin{abstract}
With the help of linear pulsation models we briefly discuss the peculiar period ratios characteristic for modulated double-mode RR Lyrae stars.
\end{abstract}

In about fifty per cent of the fundamental mode RR~Lyrae stars (RRab) we observe a quasi-periodic modulation of pulsation amplitude and phase -- the Blazhko Effect. It is also observed in single-mode first overtone stars (RRc), in which the occurrence rate is smaller. The origin of the Blazhko modulation remains a puzzle. For a review see e.g. \cite{szabo14} and Kov\'acs (these proceedings). Precise space observations significantly increased our knowledge about the Blazhko effect. The most interesting discoveries are detection of period doubling at some phases of the Blazhko cycle \citep{kol10,szabo10} and detection of additional radial modes of very low amplitude, mostly of the second overtone, and rarely of the first overtone \citep{benko10,benko14} (see also Smolec \& B\c{a}kowska, these proceedings).

\begin{figure}[!ht]
\centering
\includegraphics[width=.75\textwidth]{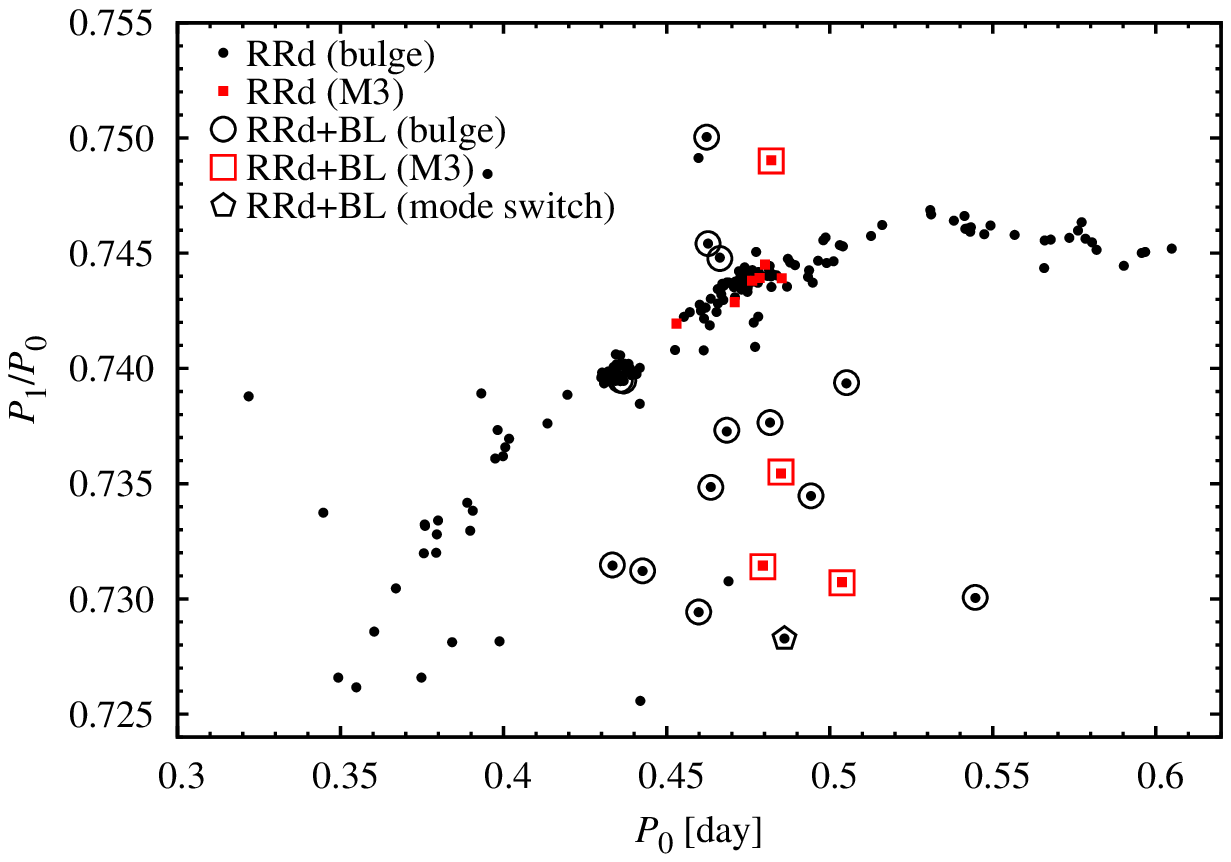}
\caption{Petersen diagram for RRd stars of the Galactic bulge \citep{ogleIV_blg,rs15a} and globular cluster M3 \citep{jurcsik_mod,jurcsik}.} 
\label{fig:pet} 
\end{figure}

Till recently modulation was not observed in double-mode RR Lyrae stars (RRd stars), pulsating simultaneously in the radial fundamental and radial first overtone modes (with large amplitude of the two). This rare form of pulsation was detected only recently in top-quality ground-based observations collected by the Optical Gravitational Lensing Experiment \citep[OGLE;][]{ogleIV} in the direction of Galactic bulge \citep[15 stars;][]{ogleIV_blg, rs15a} and in \cite{jurcsik_mod} observations of the globular cluster M3 (4 stars). The most peculiar feature of these stars is a somewhat atypical period ratio of the two radial modes. This is illustrated with the help of the Petersen diagram displayed in Fig.~\ref{fig:pet}. The majority of RRd stars in this diagram form a rather tight and curved progression; period ratio increases with the increasing fundamental mode period, up to $P_0\approx 0.52$d, and then, a mild decline is observed as period increases further. The period ratio is a well known indicator of star's metallicity; in particular the short-period tail of the RRd sequence is formed by Galactic bulge stars of relatively high metallicity \citep{ogleIII_blg}. Modulated RRd stars, except two, do not follow the described progression. Their period ratio is either too low, or too high, as compared to other RRd stars of similar fundamental mode period. Why their period ratios are atypical remains an open question.

\begin{figure}[!ht]
\centering
\includegraphics[width=.74\textwidth]{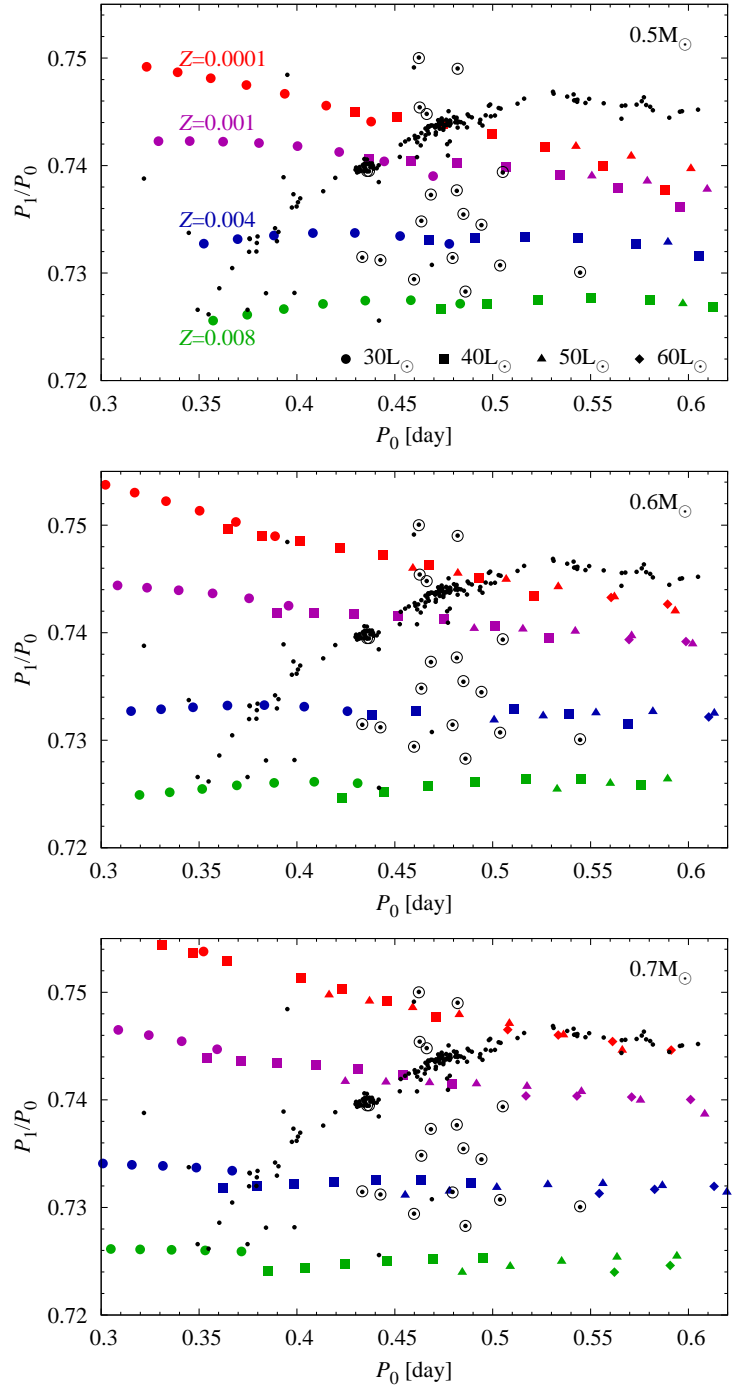}
\caption{Linear period ratios for RR Lyr models of different masses, luminosities and metallicities. RRd stars from the Galactic bulge and M3 are marked with small black dots; modulated stars are encircled.} 
\label{fig:models} 
\end{figure}

With the help of linear pulsation codes we can check whether the period ratios in modulated stars can be reproduced assuming physical parameters typical for RR Lyrae stars. Results of model calculations are presented in Fig.~\ref{fig:models}. We used pulsation codes of \cite{sm08a}, assumed three different masses, $0.5{\rm M}_\odot$, $0.6{\rm M}_\odot$ and $0.7{\rm M}_\odot$ (top, middle and bottom panels, respectively), four different luminosity levels, $30{\rm L}_\odot$ (circles), $40{\rm L}_\odot$ (squares), $50{\rm L}_\odot$ (triangles) and $60{\rm L}_\odot$ (diamonds), and a range of metallicities (plotted with different colors and labelled in Fig.~\ref{fig:models}). For a given $M$ and $L$ our model sequences are horizontal and computed with $100$K step in effective temperature. Only the models in which both fundamental mode and the first overtone are linearly unstable are plotted. This is a parameter study; parameters of our models cover the range expected for RR~Lyr stars, but these are not evolutionary models.

First we focus our attention on typical, non-modulated RRd stars that form the curved sequence in the Petersen diagram. As already mentioned, the period ratio is a good indicator of metallicity. The shortest period Galactic bulge RRd stars have higher metallicity and lower luminosities, as is expected from evolutionary calculations. The other striking feature is that models predict simultaneous instability of the two radial modes over the entire Petersen diagram. Some extreme combinations of $M/L/Z$ are likely not allowed by evolutionary calculations, but it is clear that non-modulated RRd stars occupy a much narrower band than allowed by the pulsation models. In the Hertzsprung-Russell diagram, the domain in which double-mode pulsation is present, is narrower than the domain in which the two radial modes are simultaneously unstable. This is a manifestation of a well known and difficult, non-linear problem of mode selection \citep{sm14}. Simultaneous instability of the two radial mode is only necessary, but not sufficient condition for the (non-resonant) double-mode pulsation. What are the other factors is not known. Since the majority of the non-modulated RRd stars follow the same and continuous progression in the Petersen diagram we may conclude that most likely mode selection mechanism is the same for these stars.

Now we turn to modulated RRd stars. The majority of these stars have atypical period ratio. From Fig.~\ref{fig:models} it is obvious that such period ratios present no difficulty for the pulsation theory. We observe that the lower the period ratio the higher the metallicity. Also, the lower the luminosity, the shorter the period at which the domain of simultaneous instability of the two radial modes start. For each considered mass the modulated RRd stars are easily reproduced with luminosities $\sim 40-50{\rm L}_\odot$ and a range of metallicities. It seems that there is no need to invoke any atypical masses or luminosities to explain their period ratios. Linear pulsation theory cannot help further; in particular, we cannot conclude about mode selection mechanism in these stars. This is a difficult and non-linear problem and tools we have to tackle it are insufficient \citep{sm14}.

What is the origin of the Blazhko effect in RRd stars and what is the origin of atypical period ratios remains a mystery. One of the modulated stars (marked with pentagon in Fig.~\ref{fig:pet}) switched the pulsation mode recently from RRab to RRd \citep{ogleIV_blg}. In \cite{rs15a} we hypothesize that the modulation and mode-switching may be connected. Properties of the modulated RRd stars are strongly non-stationary, amplitudes and phases of the radial modes vary on a time scale of few hundred of days. Long-term monitoring of these stars, and RR~Lyr stars in general, is needed to check whether the mode-switching and modulation are indeed connected.
\smallskip 

{\bf Acknowledgements.} This research is supported by the Polish National Science Centre through grant DEC-2012/05/B/ST9/03932.

\end{document}